\documentclass[singlespacing]{elsart}
\usepackage{amssymb}
\usepackage{ifthen, fancybox, calc, psfrag}
\usepackage{epsfig}
\usepackage{colordvi}
\begin{document}

\begin{frontmatter}

\title{A phase-field description of surface-tension-driven instability}

\author{Rodica Borcia,}
\author{Domnic Merkt}
\author{and Michael Bestehorn}
\address{Lehrstuhl f\"{u}r Theoretische Physik II, Brandenburgische Technische Universit\"{a}t Cottbus, Erich-Weinert-Stra{\ss}e 1, 03046, Cottbus, Germany}

\begin{abstract}
In this paper we report on 2D numerical simulations concerning linear and nonlinear evolution of surface-tension-driven instability in two-fluid systems heated from below using classical and phase-field models. In the phase-field formalism, one introduces an order parameter called phase-field function to characterize thermodynamically the phases. All the system parameters are assumed to vary continuously from one fluid bulk to another (as linear functions of the phase-field). The Navier-Stokes equation (with some extra terms) and the heat equation are written only once for the whole system. The evolution of the phase-field is described by the Cahn-Hilliard equation. In the sharp-interface limit the results found by the phase-field formalism recover the results given by the classical formulation.
\end{abstract}

\begin{keyword}

Instability driven by surface tension \sep  pattern formation \sep diffuse interface

\PACS 47.20.Dr \sep 47.54.+r \sep 05.70.Np
\end{keyword}
\end{frontmatter}

\section{Introduction}
\label{}

In systems with multi-material interfaces the classical methods may admit some non-physical oscillations near interfaces or meet with difficulties in the problems with large deformations. Therefore recently R. Fedkiw  \textit{et al.} developed a new numerical method$-$the so-called ``ghost-fluid'' method$-$for describing interfaces in multi-material flows [Caiden \textit{et al.}, 2001], [Fedkiw \textit{et al.}, 1999], [Fedkiw \& Liu, 2002]. In this method a so-called level set function (which satisfies the level set equation) is implemented to keep track of the interface. The zero level marks the location of the interface, the positive values correspond to one fluid and the negative values to the other. In addition they captured the appropriate interface conditions by defining a ghost fluid (for each fluid) which has the same pressure and velocity as the real fluid, but the entropy of the other real fluid. Since the ghost fluids have the same entropy as the real fluid which is not replaced, a one phase problem is solved for each fluid. Another continuum model adequate to multi-systems in which the interface location cannot be explicitly tracked is based on the phase-field method. Proposed for the first time by Langer [1986], this model was subsequently developed for studying solidification phenomena [Anderson \textit{et al.}, 2000], [Langer, 1986], [Nestler \textit{et al.}, 2000], [Wang \textit{et al.}, 1993], dendritic crystal growth [Braun \& Murray, 1997], [Tong \textit{et al.}, 2001] or dynamic fractures [Karma \textit{et al.}, 2001]. In this type of model, in place of the set level function an auxiliary variable$-$the phase-field $\varphi$ $-$is appended to the usual set of thermodynamic variables in order to provide an explicit indication of the thermodynamic phase in each point of the system. This order parameter $\varphi$ can be also regarded as a fluid concentration in a binary mixture of two fluids. The phase-field takes distinct values in each bulk phase and undergoes a rapid but smooth variation in the interfacial region. $\varphi$ is governed by a partial differential equation that guarantees the realistic interface conditions in the limit of a suitable thin interface. Therefore, the problem is treated like an entire (one phase problem), exactly as in the ghost-fluid model. Unlike the ghost fluid model, in the phase-field formalism the interface is diffuse with a finite thickness over which the fluid properties vary smoothly from one region to the other. Consequently a new phenomenon is taken into discussion: the diffusive transport in the interfacial region between the two phases.
\newline
Hence, the continuum models are suitable for describing multi-layer systems, because they reduce the system of equations (they don't need different equations for each medium) and avoid the interface conditions.

\begin{figure} 
\begin{center} 
\includegraphics[height=5.0cm, keepaspectratio=true]{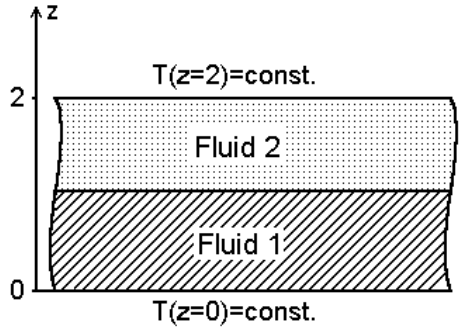}
\caption[Figure 1]{Sketch of the system: two superposed immiscible fluid layers. The system is heated from below and the temperatures on top and bottom are maintained constant.}
\end{center}
\end{figure}
\vspace*{0.5cm} 
Due to these two essential advantages, we will adjust in this paper the phase-field model to \underline{another} problem with material interface: the two-dimensional Marangoni convection (MC) induced by the temperature gradient in a two-layer system of immiscible and incompressible fluids (see Fig. 1). As it is well known, the equation which describes the bifurcation from the conductive regime to the convective regime for the system illustrated in Figure 1 can be casted into the following normal form [Colinet \textit{et al.}, 2001], [Manneville, 1990], [Newell \& Whitehead, 1969]:
\[
{{dA} \over {dt}}=\left(\beta+i\gamma\right)A-\alpha{{|A|}^2}A
\]
where $A$ denotes the amplitude measuring the "intensity" of convection and $\beta$ is the control parameter, connected with the temperature gradient. (If $Re(\alpha)$ is negative, higher order terms have to be included in the above equation in order to ensure the global stability.) When $\gamma=0$ the bifurcation from the conductive regime ($\beta<0$) to the convective regime ($\beta>0$) is realised through a monotonic instability and when $\gamma\neq0$ through an oscillatory instability. For the last situation, the Poincar\'{e} map ($dA/dt, A$) consists in a limit cycle, and this kind of bifurcations is known in literature as Hopf bifurcation [Manneville, 1990]. 
\newline 
In our paper we will report on numerical simulations concerning linear and fully nonlinear evolutions of surface-tension driven instability (or MC with short wavelength), corresponding to the two different bifurcations types mentioned above. After the introducing of the method, we shall discuss a silicon-oil$-$air system heated from below, a situation for which the short-wave instability appears as monotonic mode. Then we particularize the results for a water$-$n-octane system, heated also from below. Here the surface-tension-driven instability appears as oscillatory mode via a Hopf bifurcation. The results given by the phase-field formalism will be compared with those given by the classical model, assuming flat interface and two sets of equations for both fluids coupled by interface conditions.        

\section{The model and basic equations}
\label{}

 \begin{figure} 
\begin{center}
\psfrag{t}{\textbf{\scriptsize {$\varphi$}}}
\psfrag{q}{\textbf{\tiny {$Z$}}}
\includegraphics[height=4.0cm, keepaspectratio=true, angle=270]{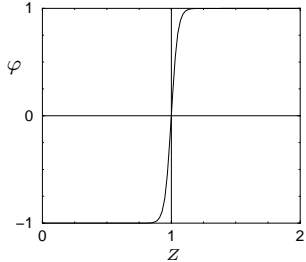}
\caption[Figure 2]{Phase-field distribution versus $z$ for the stationary state. At $z=0$ (fluid 1 boundary) the phase-field function takes the value $\varphi=-1$ and at $z=2$ (fluid 2 boundary) the value $\varphi=+1$. The diffuse interface between the two different immiscible fluids is situated around $z=1$.}
\end{center}
\end{figure}

The phase-field is assumed to be $\varphi=-1$ at the lower boundary of fluid 1 ($z=0$) and $\varphi=+1$ at the upper boundary of fluid 2 ($z=2$) while the diffuse interface between the two different immiscible fluids is situated around $z=1$ for the simulations examined in the present paper (Fig.2). The system is bounded in the vertical direction by two solid, perfectly heat conducting walls with fixed temperatures $T(z=0)=T_b$ and $T(z=2)=T_t$. In the phase-field formalism the Navier-Stokes (NS) equation must contain some non-classical phase-field terms. In the following we shall derive these supplementary terms by minimizing the free-energy functional for our system [Anderson \textit{et al.}, 1998], [Jasnow \& Vi\~nals, 1996], [Langer, 1986]:

\begin{figure} 
\begin{center}
\psfrag{g}{\textbf{\scriptsize {$f(\varphi)$}}}
\psfrag{p}{\textbf{\scriptsize {$\varphi$}}}
\includegraphics[height=4.0cm, keepaspectratio=true, angle=270]{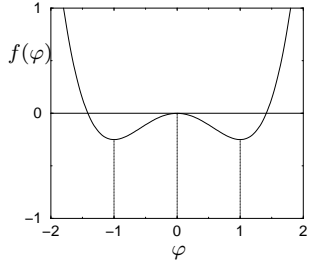}
\caption[Figure 3]{Free-energy density representation versus phase-field function $\varphi$. We have chosen $f(\varphi)$ as a symmetric "double-well" potential with two local minima: one at $\varphi=-1$, corresponding to the fluid 1, and the second at $\varphi=+1$, corresponding to the fluid 2.}
\end{center}
\end{figure}

\begin{equation}
\mathcal{F}[\varphi]=\int_V\left[f(\varphi)+{\mathcal{K}(T) \over 2}(\nabla\varphi)^2\right]dV.
\end{equation}

The first term from the above functional represents the free-energy density for an homogeneous system corresponding to regions far from interphase boundaries. The latter one characterizes the energy concentrated in the interfacial region. $f(\varphi)$ is considered to be given by the expression $f(\varphi)=C({\varphi^{4}\over4}-{\varphi^{2}\over2})$ ($C-$positive constant) [Borcia \& Bestehorn, 2003] which is a symmetrical curve with two local minima: one corresponding to $\varphi=-1$, namely for the fluid 1 bulk, and another one to $\varphi=+1$, for fluid 2 (see Fig.3). The coefficient $\mathcal{K}$ in (1) is connected to the surface tension coefficient, because the second term from the functional (1) is associated with variations of the density (and consequently of the phase-field) and contributes to the free-energy excess of the interface [Jasnow \& Vi\~nals, 1996], [Rowlinson \& Widom, 1982]:

\begin{equation}
\sigma=\int_{-\infty}^{+\infty} \mathcal{K}{\left({\partial \varphi \over \partial z}\right)^2}dz
\end{equation}
In many previous works $\mathcal{K}$ is assumed to be constant [Anderson \textit{et al.}, 1998], [Jasnow \& Vi\~nals, 1996], [Langer, 1986]. But the short-wave instability is driven by the gradients of the surface-tension. To include the thermo-capillary effects it is necessary to consider $\mathcal{K}$ weakly dependent on temperature:
\begin{equation}
\mathcal{K}=\mathcal{K}_0-{\mathcal{K}_T}T \hspace{10pt}(\mathcal{K}_T>0).
\end{equation}  
The Lagrangian energy density $\mathcal{L}$ of our system which is assumed to have an explicit dependence on the scalar $\varphi$, its gradient $\nabla\varphi$ and on the gradient energy coefficient $\mathcal{K}(T)$
\begin{equation} 
\mathcal{L}\left(\varphi, \nabla\varphi, \mathcal{K}(T)\right) =f(\varphi)+{\mathcal{K}(T) \over 2}(\nabla\varphi)^2
\end{equation}
must satisfy the Euler-Lagrange equation:
\begin{eqnarray}
{\partial \mathcal{L} \over \partial\varphi}-\sum_{i=1}^{3} {\partial \over \partial x_i} \left(\partial\mathcal{L} \over \partial\left(\partial_i \varphi\right)\right)=0.
\end{eqnarray}
Substitution of relation (4) in (5) gives us the following relation (Noether's theorem):
\begin{equation}
\sum_{i, j=1}^{3} \left[ {\partial \over \partial x_i} \left( {{\partial \mathcal{L} \over \partial \left(\partial_i \varphi \right)} {\partial\varphi \over \partial x_j} - \mathcal{L} \delta_{ij}}\right)+{\partial\mathcal{L} \over \partial\mathcal{K}}{\partial\mathcal{K} \over \partial x_i}\delta_{ij}\right]=0.
\end{equation}
Assuming $\mathcal{K}=const.$, the stress tensor is defined as: 
\[
\Xi=\mathcal{K}\nabla\varphi\otimes\nabla\varphi-\mathcal{L}\mathcal\ {I},
\] 
or, in components,
\[
\Xi_{ij}=\mathcal{K}{\partial\varphi \over \partial x_i}{\partial\varphi \over \partial x_j}-\mathcal{L}\delta_{ij},
\]
and (6) can be considered as a conservation law:
\begin{equation}
\nabla\cdot\Xi=0.
\end{equation}
Under these conditions, a conservative term $-\nabla\cdot\Xi$ appears in the Navier-Stokes equation. This term describes the contribution of capillary forces [Anderson \textit{et al.}, 2000]:
\begin{eqnarray}
\rho{d\vec{v} \over dt}&=&\nabla\left(-p+\mathcal{L}\right)-\nabla\cdot(\mathcal{K}\nabla\varphi\otimes\nabla\varphi)\;\nonumber\\
& &+\nabla\cdot(\eta\nabla\vec{v})%
      +\rho g\hat{z}. 
\end{eqnarray}
(Both fluids are assumed incompressible.)
But if $\mathcal{K}\neq const.$, the conservative law (7) is not satisfied. Hence we must consider also the last term of relation (6). Therefore a new force-component has to be included:
\begin{eqnarray}
\rho{d\vec{v} \over dt}&=&\nabla\left(-p+\mathcal{L}\right)-\nabla\cdot(\mathcal{K}\nabla\varphi\otimes\nabla\varphi)\;\nonumber\\
& &+\nabla\cdot(\eta\nabla\vec{v})+\rho g\hat{z}+{1\over2}{\mathcal{K}_T}{\nabla T}{(\nabla\varphi)^2}. 
\end{eqnarray}
This last term from (9) represents the ``Marangoni force'' which is the trigger for the short-wavelength instability, because integrating equation (9) through the interface, this new-force component assures the fulfilment of the interface condition for tangential stresses in the limit of sharp and rigid interfaces [Borcia \& Bestehorn, 2003]:
\[
{\sigma'_{xz}}(2)-{\sigma'_{xz}}(1)={\partial \sigma \over \partial x},
\]
($\sigma'_{xz}=\eta\left({{\partial v_x \over \partial z}+{\partial v_z \over \partial x}}\right)$ is the viscous stress tensor for incompressible fluids).
Equation (9) together with the energy equation 
\begin{equation}
\rho c{dT\over dt}=\nabla\cdot(\kappa\nabla T),
\end{equation}
($\rho$ is the fluid density, $c-$the heat capacity and $\kappa-$the thermal conductivity) and the Cahn-Hilliard (CH) equation, which serves for mass conservation in a system with diffuse interface
\begin{equation}
\dot{\varphi}=-M_o \triangle \left( \mathcal{K} \triangle \varphi-{\partial f \over \partial \varphi} \right)
\end{equation} 
($M_o-$the phenomenological mobility), describe both Marangoni instabilities without to impose supplementary interface conditions. In the classical models the basic equations are written two times: once for the fluid 1 and the second time for the fluid 2 and matched by interface conditions. In our phase-field formalism the NS equation (9), the heat equation and the CH equation are valid for both fluids while all the system parameters are assumed to be linear functions of the phase-field:
\[
\rho={{\rho_1+\rho_2} \over 2}-{{\rho_1-\rho_2} \over 2}\varphi, \hspace{10pt}
\eta={{\eta_1+\eta_2} \over 2}-{{\eta_1-\eta_2} \over 2}\varphi
\]
\[
\hspace{5pt}
c={{c_1+c_2} \over 2}-{{c_1-c_2} \over 2}\varphi, \hspace{10pt}
\kappa={{\kappa_1+\kappa_2} \over 2}-{{\kappa_1-\kappa_2} \over 2}\varphi.
\]
In the above relations the index ``1'' describes the fluid parameters at $z=0$, while the index ``2'' describes the fluid parameters at $z=2$.
After adimensionalization of (9)-(11)
\[
\vec{r}=d_1\vec{r'},\hspace{10pt} t={{d_1^2} \over \chi_1}t',\hspace{10pt}  \vec{v}={\chi_1 \over d_1}\vec{v'},\hspace{10pt}  T'={{T-T_t} \over {T_b-T_t}},
\]
\[ \hspace{10pt} \rho=\rho_1 \rho', \hspace{10pt} \eta=\eta_1 \eta', \hspace{10pt} c=c_1 c', \hspace{10pt} \kappa=\kappa_1 \kappa' 
\]
($\chi_1-$thermal diffusivity at the lower boundary of fluid 1, $d_1-$the fluid 1 depth)
 the following non-dimensional parameters appear:
\newline
$P_r={\eta_1 \over {\rho_1 \chi_1}}-$Prandtl number of the fluid 1,
\newline 
$Ca={{\sqrt{\mathcal{K}_0 C}d_1} \over {\eta_1\chi_1}}-$capillary number,
\newline 
$M={{{\mathcal{K}_T\sqrt{\mathcal{K}_0 C}} \over {\mathcal{K}_0}}{{(T_b-T_t)d_1} \over {\eta_1\chi_1}}}-$Marangoni number,
\newline
 $G={{\rho_1 gd_1^3}\over {\eta_1\chi_1}}-$Galileo number,
\newline 
$L={1 \over d_1}\sqrt{{\mathcal{K}_0 \over C}}-$the width of interface
\newline
$M_o'={M_oC \over {\chi_1}}-$the phenomenological mobility.
\newline
Assuming both fluids being incompressible, the velocity field, $\vec{v}(x,z,t)$  can be replaced by the stream-function, $\psi(x,z,t)$ (the primes are dropped):
\[
\vec{v}={\partial \psi \over \partial z}\vec{i}-{\partial \psi \over \partial x}\vec{k},
\]
with $\vec{i}$ and $\vec{k}$ the unit vectors in $x$ and $z$ direction, respectively.

\begin{figure} 
\begin{center}
\includegraphics[height=7.0cm, keepaspectratio=true]{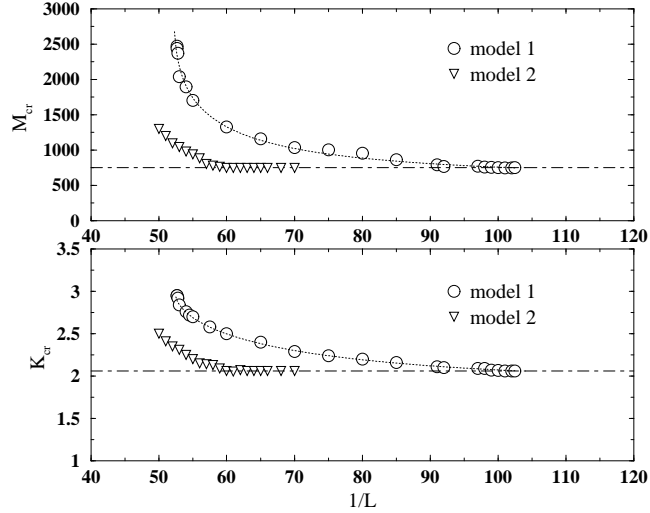}
\caption[Figure 4]{Dependencies of critical Marangoni number $M_{cr}$ and critical wavenumber $k_{cr}$ on the interface thickness, for MC driven by surface tension gradient. The plots correspond to two different models: model 1 assumes a perfect rigid liquid-gas interface, while in model 2 the interface is quasi non-deformable. For model 2 one considers: $C_a=2\cdot10^5, G=3\cdot10^9$. Both models converge in the limit of sharp interfaces to classical results obtained for systems with flat interface.}
\end{center}
\end{figure}
\vspace*{0.5cm}

\section{Purpose}
\label{}

 For small capillary numbers $Ca$ and Galileo numbers $G$, that is, for very thin layers, the interface between the two immiscible fluids is deformable. Thereby the long-wave instability induced by surface deflections appears even for small Marangoni numbers. For large capillary and Galileo numbers the interface becomes quasi-non-deformable and in the system develops only one instability: the short-wave instability. The goal of our paper is to analyze this instability in the frame of the phase-field model for the following situations:

1. the interface is perfectly rigid but diffuse (model 1); in this case we drop CH equation (11) and we assume for the phase-field in (9) a particular variation of the form:
\[
\varphi^{(0)}(z)=\tanh{\left({z-1} \over {L \sqrt{2}}\right)}.
\]

2. the interface is quasi-non-deformable; for this situation the full system of equations (9)-(11) is considered, but in the limit of large values for capillary and Galileo numbers (model 2).

The results described by these two situations will be compared with the results given by the classical model, when the NS equation and the heat equation are written for both media [Engel \& Swift, 2000]. At $z=1$ the interface conditions are given by [Colinet \textit{et al.}, 2001], [Engel \& Swift, 2000]:
\[
\psi_1=\psi_2=0, \hspace{10pt} {\partial \psi_1 \over \partial z}={\partial \psi_2 \over \partial z}, \hspace{10pt} \Theta_1=\Theta_2, \hspace{10pt}  \kappa_1{\partial \Theta_1 \over \partial z}=\kappa_2{\partial \Theta_2 \over \partial z}, 
\]
\[
{\partial^2 \psi_1 \over \partial z^2}-{{\rho_2 \nu_2} \over {\rho_1 \nu_1}}{\partial^2 \psi_2 \over \partial z^2}=M'{\partial^2 \Theta \over \partial x^2},
\]
where the subscripts $1(2)$ denotes the lower (upper) fluid, $\Theta$ represents the temperature perturbation, $\nu-$the kinematic viscosity and $M'$ is the usual definition of the Marangoni number [Engel \& Swift, 2000]:

\begin{figure} 
\begin{center}
\includegraphics[height=6.0cm, keepaspectratio=true, angle=270]{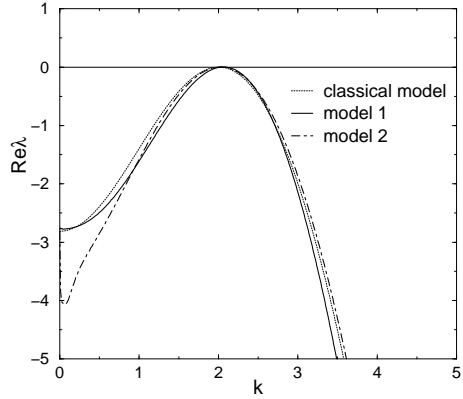}
\caption[Figure 5]{Growth rate of surface-tension-driven-instability, $Re \hspace{2pt} \lambda$ versus $k$ near threshold as it results from classical and phase-field models. The three curves agree well near $k=2$.}
\end{center}
\end{figure}
\vspace*{0.5cm}

\begin{equation}
M'=M{B_i\over {B_i+1}},
\end{equation}
(with $B_i=\kappa' /a-$the Biot number, $\kappa'=\kappa_{2}/\kappa_{1}$ and $a=d_{2}/d_{1}$).

\section{Results}
\label{}

Results concerning the linear stability analysis for the three models in bi-dimensional (x, z) representation are emphasized in Figs. 4 and 5 for a $10cS$ silicon-oil$-$air system ($\rho_1=940 \hspace{3pt} kg/m^3$, $\rho_2=1.2 \hspace{3pt} kg/m^3$, $\nu_1=0.1 \hspace{3pt} cm^2/s$, $\nu_2=0.015 \hspace{3pt} cm^2/s$, $\kappa_1=0.142 \hspace{3pt} J/msK$, $\kappa_2=0.026 \hspace{3pt} J/msK$, $c_{p1}=1450 \hspace{3pt} J/kgK$, $c_{p2}=1042 \hspace{3pt} J/kgK$ [Golovin \textit{et al.}, 1997]) heated from below with $a=1$. For this stability analysis the small perturbations are assumed as
\[
\left( \begin{array}{c} \psi(x,z,t) \\ \varphi(x,z,t) \\ T(x,z,t)\\  \end{array} \right)=\left( \begin{array}{c} \psi^{(0)}(z) \\ \varphi^{(0)}(z) \\ T^{(0)}(z)\\  \end{array} \right)+\left( \begin{array}{c} \psi^{(1)}(z) \\ \varphi^{(1)}(z) \\ T^{(1)}(z)\\  \end{array} \right)exp(ikx)exp(\lambda t)
\]
with $k-$wavenumber and $\lambda-$the (complex) growth-rate. The derivatives with respect to $z$ are expressed using a finite-difference method.
Figure 4 represents the critical Marangoni number and the critical wavenumber versus $1/L$ for the two models 1 and 2. One can see how in the limit of sharp interfaces the results given by the phase-field models saturate for large values of $1/L$. The critical Marangoni number saturates around $M_{cr}=750$, which means in the terms of the usual definition (12) a value around $M'_{cr}=110$.

\begin{figure}
\begin{center}
\includegraphics[height=16.0cm, keepaspectratio=true]{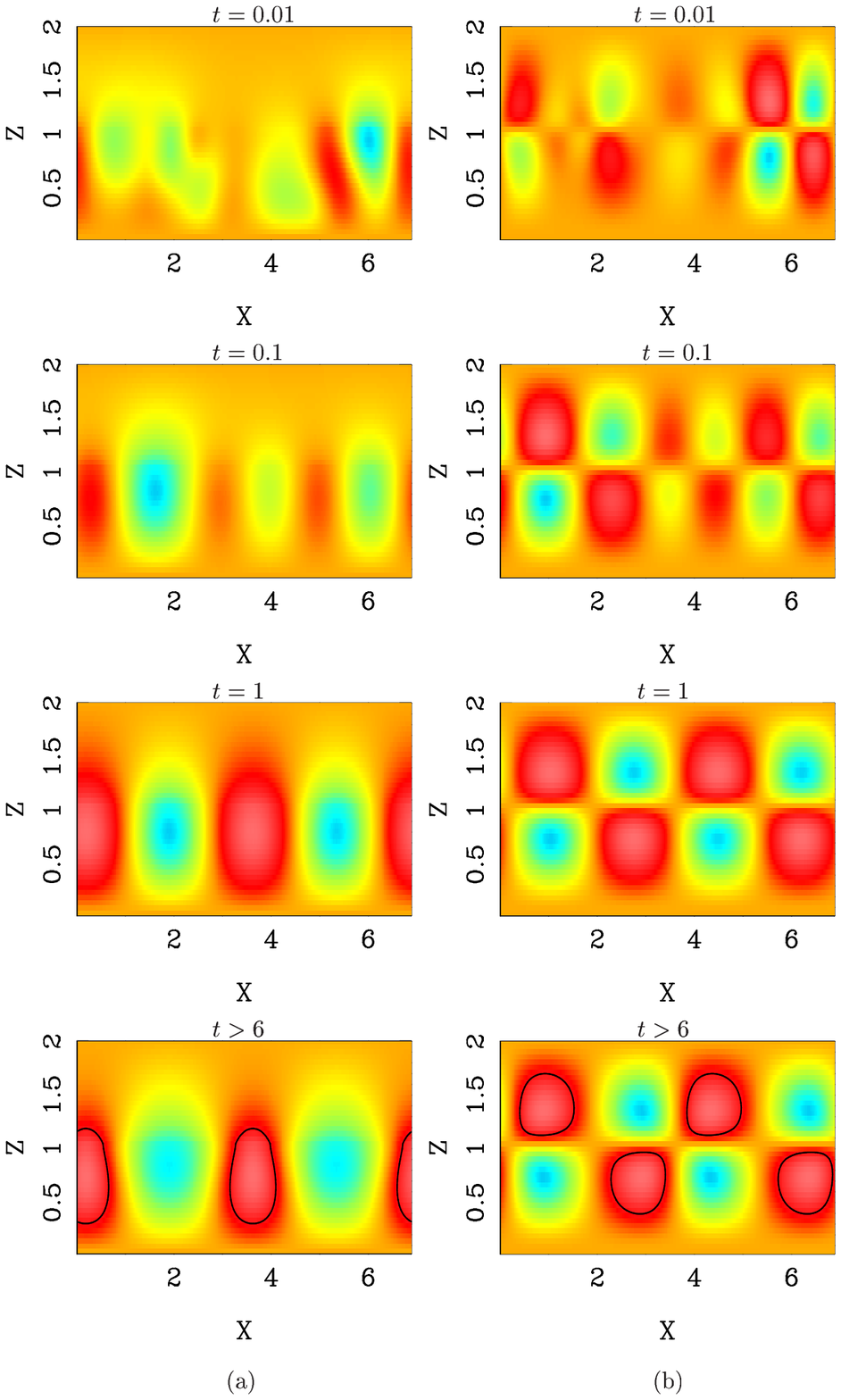}
\caption[Figure 6]{Time series for the perturbation profiles in the surface-tension-driven case as it results from the classical formalism. The panels (a) correspond to the temperature perturbations and the panels (b) to the stream-function perturbations. Starting from an initial random pattern the perturbation evolution is represented till the saturation state.}
\end{center}
\end{figure}
\vspace*{0.5cm}

\begin{figure}
\begin{center}
\includegraphics[height=16.0cm, keepaspectratio=true]{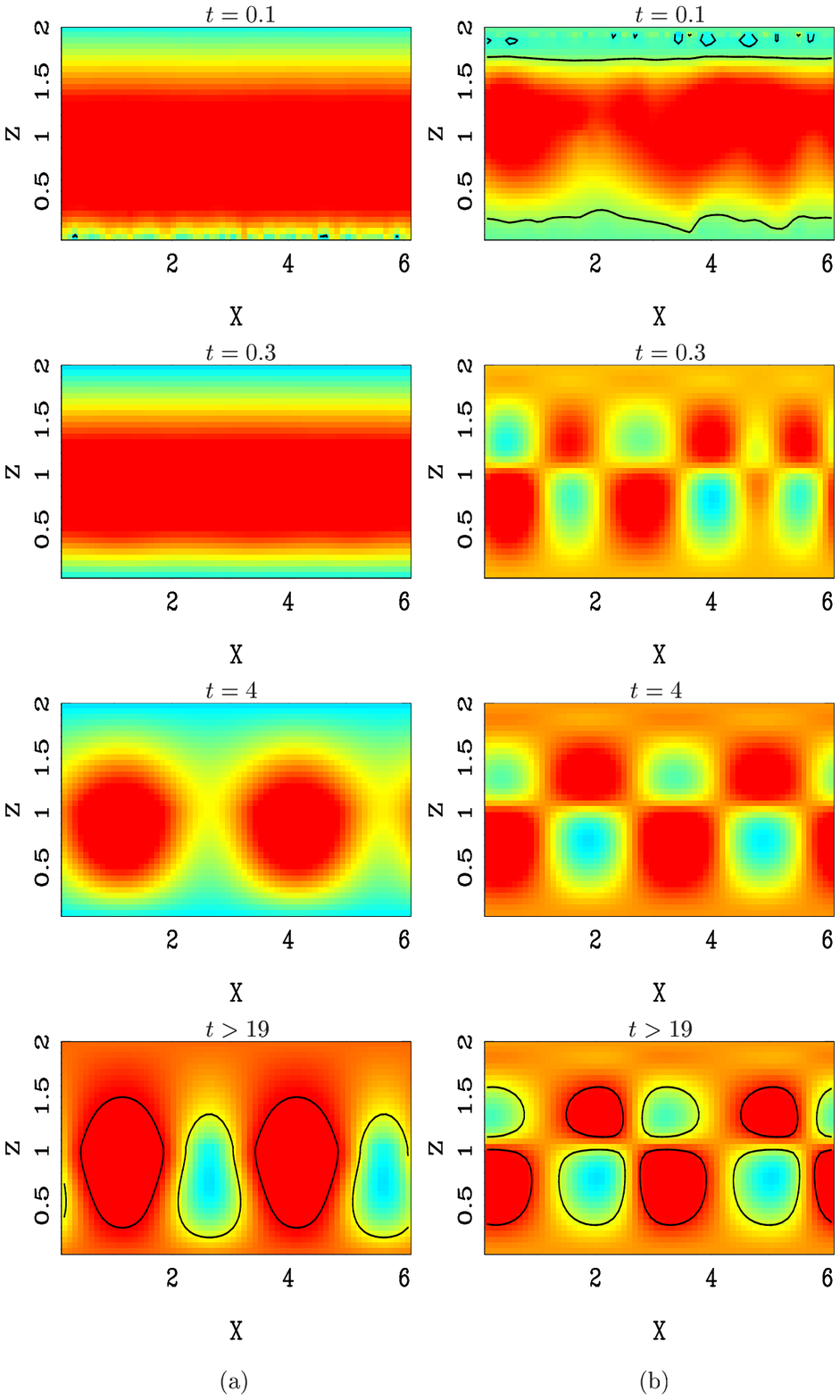}
\caption[Figure 7]{Same as Fig. 6, but in the phase-field description with rigid interface (model 1).}
\end{center}
\end{figure}
\vspace*{0.5cm}

\begin{figure}
\begin{center}
\includegraphics[height=15.0cm, keepaspectratio=true]{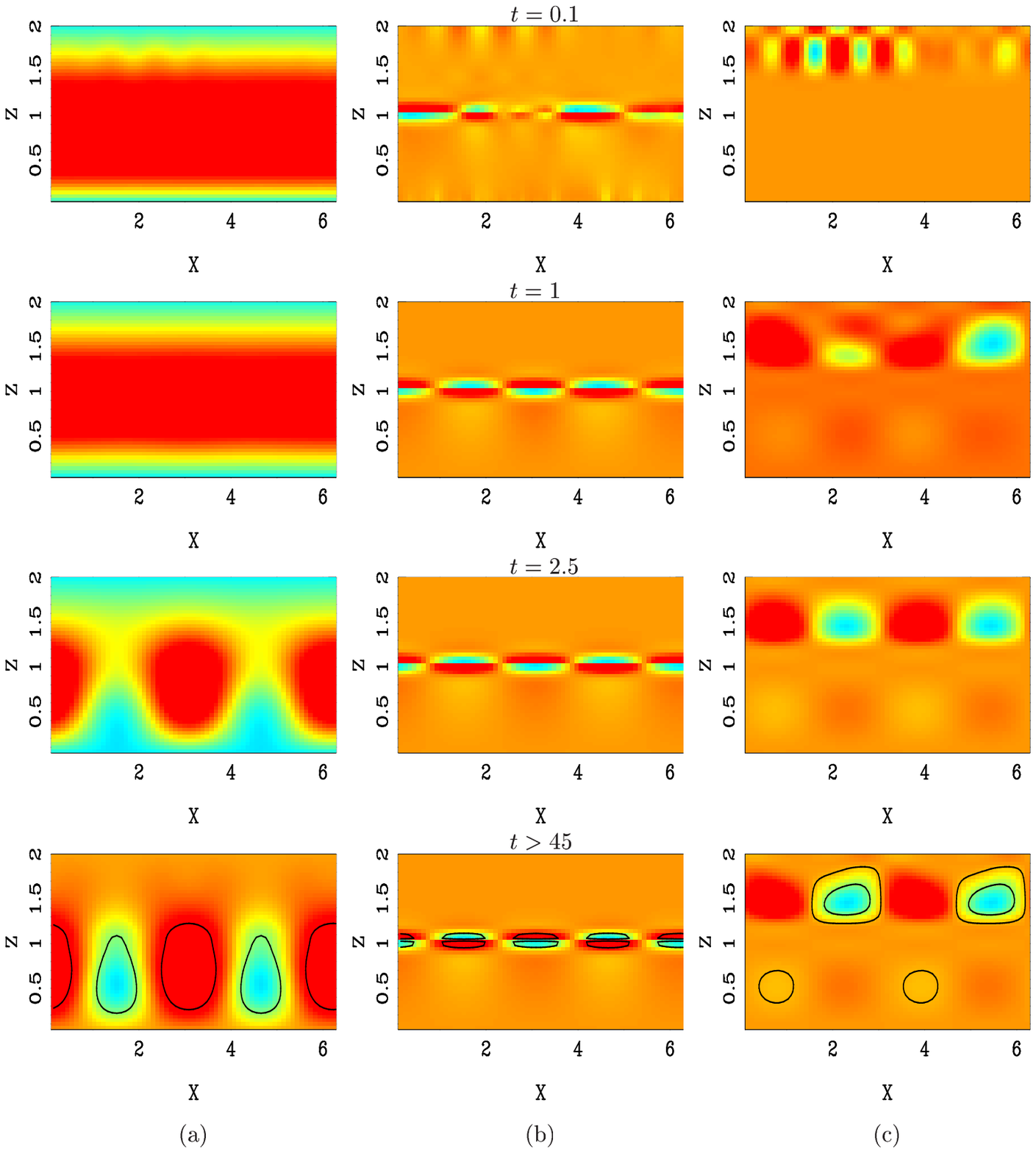}
\caption[Figure 8]{Same as Fig. 6, but for model 2, when the liquid-gas interface is quasi-non-deformable. The panels (a) display the perturbation profile of the temperature, panels (b) of the phase-field and panels (c) of the stream-function.}
\end{center}
\end{figure}
\vspace*{0.5cm}

This value is in good agreement with the critical Marangoni number obtained by the classical model for the same $10cS$ silicon-oil$-$air combination with $a=1$: $M'_{cr}\cong 93$. If one compares the critical wavenumber given by the models 1 and 2 in the sharp-interface limit with the critical wavenumber given by the classical model, the agreement is much better. For model 1 one obtains: $k_{cr}=2.06$, for model 2: $k_{cr}=2.04$ and from the classical model one obtains for the critical wavenumeber $k_{cr}=2.003$. Figure 5 shows the growth-rate of Marangoni instability with short-wavelength at threshold versus the wavenumber $k$ as it results from classical and non-classical models. One can see a good concordance between the curves obtained with the classical model and those obtained with the phase-field model for sharp interfaces, especially around the threshold. For model 2 the interface between the liquid and the gas is quasi-non-deformable, but not completely rigid. For this situation, the curve $Re\lambda=f(k)$ indicates the second Marangoni instability, the long-wave instability, which develops around $k=0$ in systems with a deformable interface.
  
In order to describe the nonlinear evolution of short-wave instability for both diffusive models and for the classical model, the corresponding systems of equations were numerically solved. We used a semi-implicit scheme: implicitly for the linear part and explicitly for the nonlinear one. A fast Fourier transform was applied in $x$ direction and the derivatives with respect to $z$ were expressed in finite-differences. To avoid the convolution sums in Fourier space, the nonlinearities are calculated in real space using again a finite difference method with grid spacing $\Delta x$ and $\Delta z$ for their derivatives. Figures 6, 7 and 8 display bi-dimensional $(x,z)$ representations of the system perturbations for each model for the same oil$-$air combination and for the same $\varepsilon$
\[
\varepsilon={{M-M_{cr}} \over M_{cr}},
\]
$\varepsilon$=0.2. All the plots in the following concerning phase-field model correspond to large values for $1/L$. Starting from an initial random distribution of very small amplitude, the pattern evolution is shown till saturation (moment emphasized by the contour lines). The time series were calculated on a mesh with $256\times40$ points for the classical model, $64\times37$ points for model 1 and $64\times51$ points for model 2. For the frames depicted in Figs. 7 and 8 (corresponding to the models 1 and 2) the integration time step is $\Delta t=0.1$, and for those from Fig. 6 (classical model) $\Delta t=10^{-4}$. Choosing an iteration time step $\Delta t=0.1$ for the classical model, one obtains very quickly (after a few iteration) the specific pattern for short-wave instability (two-convective motions developed in the liquid and in the gas). The presence of a diffusive interface decelerates the process of pattern formation for the short-wavelength instability and also the process of saturation compared to the classical case where the interface is well defined. If additionally the diffuse interface becomes weakly deformable, the processes of pattern formation and saturation become longer. For the latter situation one can also follow the time evolution for the phase-field which in fact describes the time evolution for the small deflections of the interface (see Fig. 8-b). These small deformations of the interface affect the aspects of the convective motions. In this case the  deflections allows a momentum transfer from the liquid to the gas and vice-versa. Because the viscosity in the liquid is one order of magnitude larger than the viscosity in the gas, it follows that the convection in the gas becomes stronger in comparison to the convection in the liquid as indicated in Fig. 8-c.
For all the situations represented in Figs. 6, 7 and 8 the short-wave instability appears as monotonic instability.

\begin{figure}
\begin{center}
\includegraphics[height=16.0cm, keepaspectratio=true]{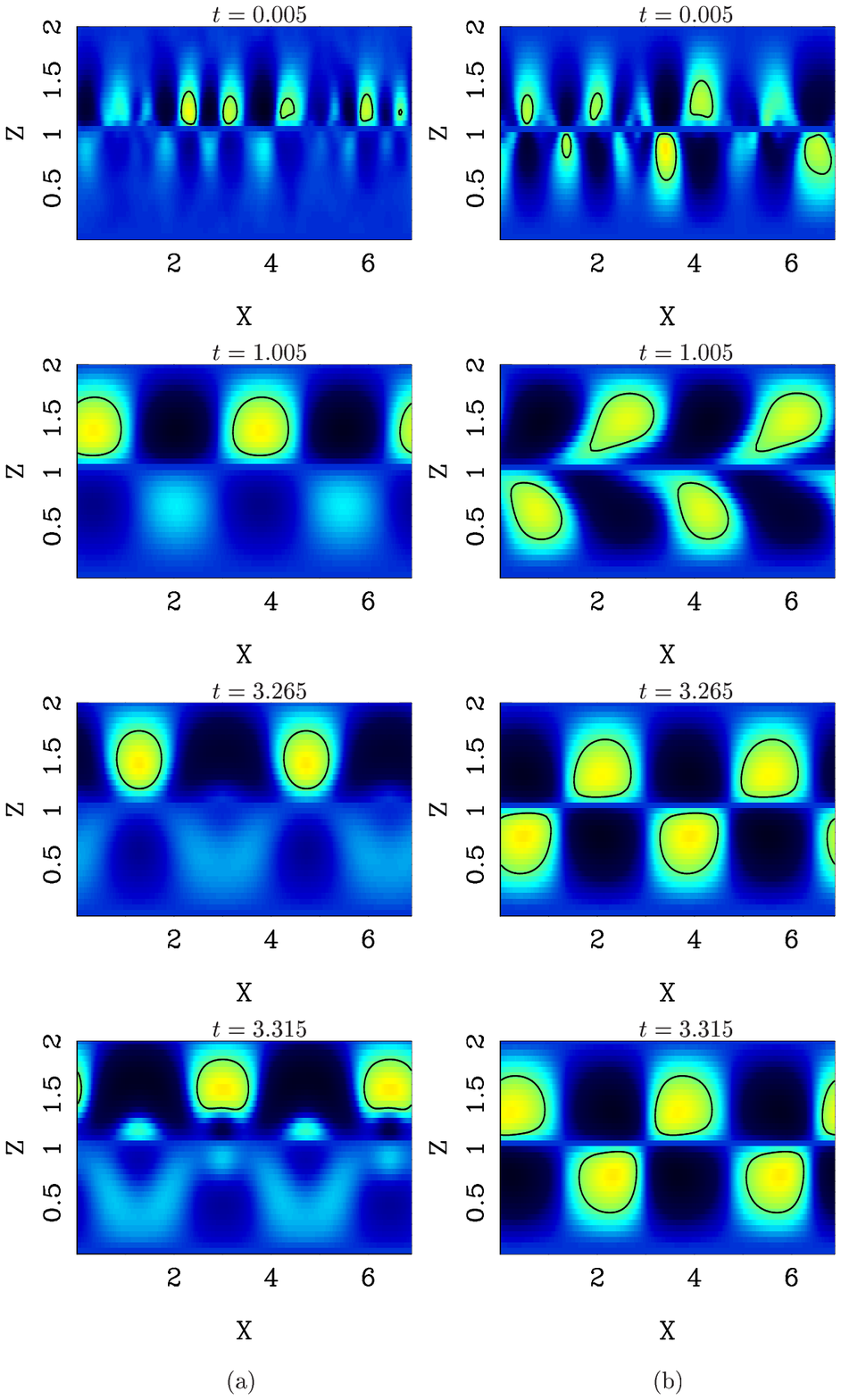}
\caption[Figure 9]{Time series for perturbations corresponding to short-wave oscillatory instability as it results from the classical formulation. The left panels (a) describe the temperature perturbation and the right ones (b) the stream-function perturbation. Starting again from a random initial distribution the first two sequences depict the formation of short-wave instability. The last two sequences show the evolution during one complete period when saturation is already reached.}
\end{center}
\end{figure}
\vspace*{0.5cm}

\begin{figure}
\begin{center}
\includegraphics[height=16.0cm, keepaspectratio=true]{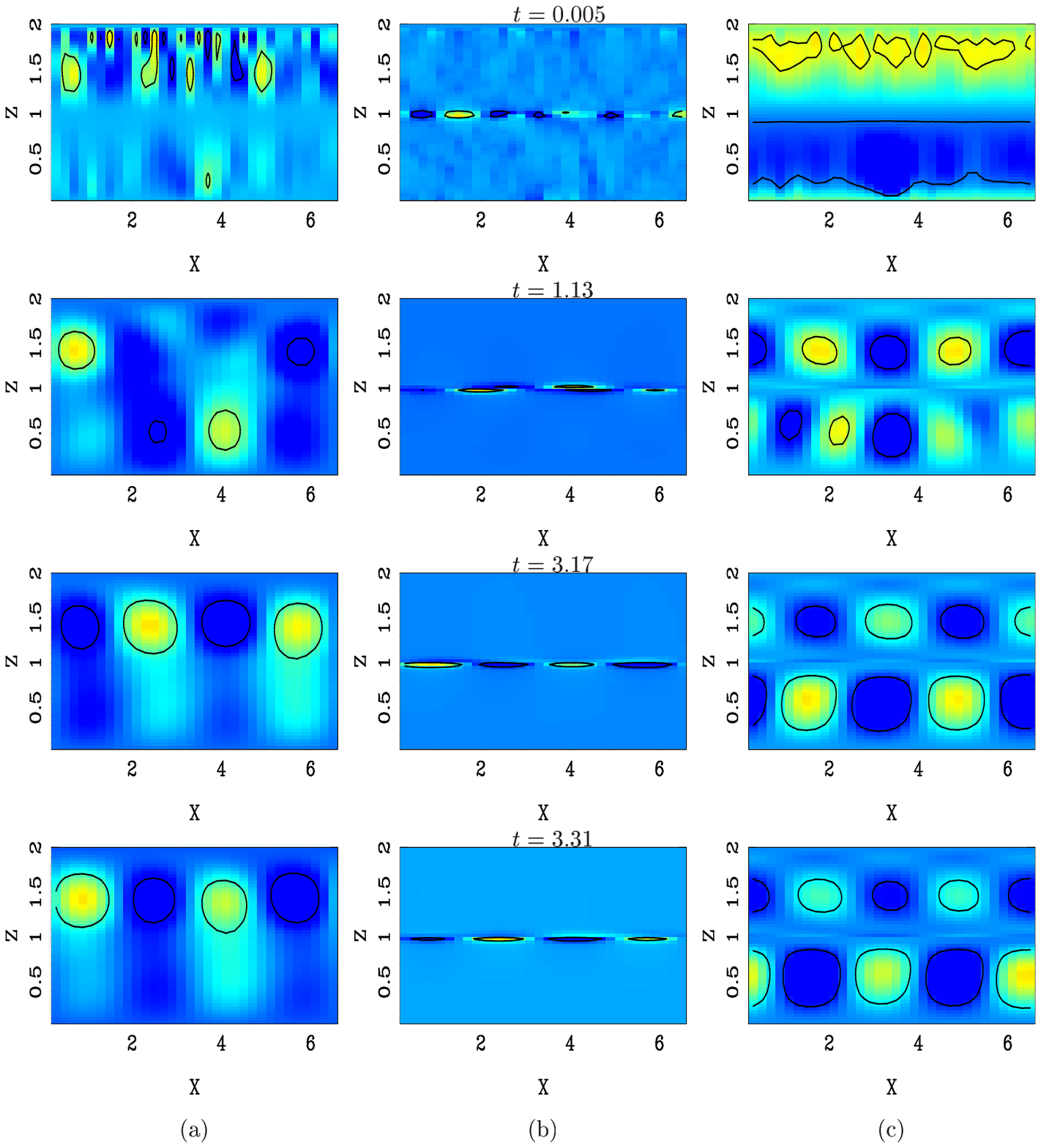}
\caption[Figure 10]{Same as Fig. 9, but in phase-field description with quasi non-deformable interface (model 2): (a) temperature, (b) phase-field and (c) stream-function.}
\end{center}
\end{figure}
\vspace*{0.5cm}

An oscillatory instability (corresponding to a Hopf bifurcation) appears in a water$-$n-octane system ($\rho_1=998 \hspace{3pt} kg/m^3$, $\rho_2=704 \hspace{3pt} kg/m^3$, $\nu_1=0.01 \hspace{3pt} cm^2/s$, $\nu_2=0.00813 \hspace{3pt} cm^2/s$, $\kappa_1=0.6 \hspace{3pt} J/msK$, $\kappa_2=0.15 \hspace{3pt} J/msK$, $c_{p1}=4182 \hspace{3pt} J/kgK$, $c_{p2}=2091 \hspace{3pt} J/kgK$ [Colinet \textit{et al.}, 2001]) heated from below with $a=1$. In this case the critical Marangoni number is: $M_{cr}\cong1.5\cdot 10^5$ and the critical wavenumber: $k_{cr}\cong1.9$. With these parameters and $\varepsilon=0.2$ we have illustrated in Figs. 9 and 10 some snapshots till saturation state is achieved. The frames depicted in Fig. 9 correspond to the classical model (with a mesh of $256\times40$ points) and those from Fig. 10 to the model 2 (with a mesh of $32\times51$ points). In the saturated state the patterns for both models are standing waves. More generally, model 2 is able to reproduce the time evolution of the stream-function and temperature perturbations described by the classical formalism; moreover one can follow the interface dynamics during the oscillation processes, a big advantage of the phase-field model.

\section{Conclusions}
\label{}

We developed a phase-field model for Marangoni convection in a two-layer-system with diffuse interface. We reported on 2D simulations concerning surface-tension-driven instability via phase-field and via classical models. Linear and nonlinear behavior for this kind of instability is examined. In contrast to the classical formulation, the phase-field model treats the problem continuously, no interface needs to be tracked. In the sharp-interface limit our results emphasize how the phase-field model applied on MC recover the results given by two-layer-systems with interface conditions. Using the same phase-field formalism, we plan to extend our research to the long-wave instability induced by surface deflections and to the influence of evaporation on Marangoni convection, problems which can be discussed in this approach in a more natural way.

\section*{Acknowledgments}

One of the authors (R.B.) acknowledges financial support from European Union under the network ICOPAC (Interfacial Convection and Phase Change) HPRN-CT-2000-00136.


\begin{thebibliography}{00}

\bibitem{And1} Anderson, D. M., McFadden G. B. \& Wheeler, A. A. [1998], ``Diffusive-interface methods in fluid mechanics'', \textit{Annu. Rev. Fluid Mech.} \textbf{30}, 139-165.
\bibitem{And2} Anderson, D. M., McFadden, G. B. \& Wheeler, A. A. [2000] ''A phase-field model of solidification with convection'', \textit{Physica D} \textbf{135}, 175-194.
\bibitem{Bor} Borcia, R. \& Bestehorn, M. [2003] ``Phase-field model for Marangoni convection in liquid-gas systems with deformable interface'', \textit{Phys. Rev. E} \textbf{67}, 066307-1-10.
\bibitem{Braun} Braun, R. J. \& Murray, B. T. [1997] ``Adaptive phase-field computations of dendritic crystal growth'', \textit{J. Cryst. Growth} \textbf{174}, 41-53.
\bibitem{Cai} Caiden, R., Fedkiw, R. \& Anderson, C. [2001] ``A numerical method for two phase flow consisting of separate compressible and incompressible regions'', \textit{J. Comput. Phys.} \textbf{166}, 1-27.
\bibitem{Col} Colinet, P., Legros, J. C. \& Velarde, M. G. [2001] \textit{Nonlinear Dynamics of Surface-Tension-Driven Instabilities} (Wiley-V. C. H., Berlin) pp. 93-154.
\bibitem{Engel} Engel, A. \& Swift, J. B. [2000] ``Planform selection in two-layer B\'enard-Marangoni convection'', \textit{Phys. Rev. E} \textbf{62}, 6540-6553.
\bibitem{Fed} Fedkiw, R., Aslam, T., Merriman, B. \& Osher, S. [1999] ``A non-oscillatory Eulerian approach to interfaces in multimaterial flows (the Ghost Fluid Method)'', \textit{J. Comput. Phys.} \textbf{152}, 457-492.
\bibitem{Fed1} Fedkiw R. \&  Liu, X.-D. [2002] ``The Ghost Fluid for viscous flows'' in \textit{Innovative methods for numerical solutions of partial differential equations}, eds. Hafez, M. \& Chattot, J.-J. (World Scientific Publishing, New Jersey) pp. 111-143.
\bibitem{Gol} Golovin, A. A., Nepomnyashchy, A. A. \& Pismen, L. M. [1997] ``Nonlinear evolution and secondary instabilities of Marangoni convection in a liquid-gas system with deformable interface'', \textit{J. Fluid Mech.} \textbf{341}, 317-341.
\bibitem{Jas} Jasnow, D. \& Vi\~nals, J. [1996] ``Coarse-grained description of thermo-capillary flow'', \textit{Phys. Fluids} \textbf{8}, 660-669.
\bibitem{Karma} Karma, A., Kessler, D. A. \& Levine, H. [2001] ``Phase-field model of mode III dynamic fracture'', \textit{Phys. Rev. Lett.} \textbf{87}, 045501-1-4.
\bibitem{Lang} Langer, J. S. [1986] ``Models of pattern formation in first-order phase transitions'', in \textit{Directions in Condensed Matter}, eds. Grinstein, G. \& Mazenko, G. (World Scinetific, Singapore) pp. 165-186.
\bibitem{Mann} Manneville, P. [1990] \textit{Dissipative structures and weak turbulence} (Academic Press, San Diego) Chap.5 pp. 137-182.
\bibitem{Nes} Nestler, B., Wheeler, A. A., Ratke, L. \& Stocker, C. [2000] ``Phase-field model for solidification of a monotectic alloy with convection'', \textit{Physica D} \textbf{141}, 133-154.
\bibitem{New} Newell, A.C. \& Whitehead, J.A. [1969] ``Finite bandwidth, finite amplitude convection'', \textit{J. Fluid. Mech.} \textbf{38}, 279-303. 
\bibitem{Row} Rowlinson, J. S. \& Widom, B. [1982] \textit{Molecular Theory of Capillarity} (Clarendon Press, Oxford) pp. 56.
\bibitem{Tong} Tong, X., Beckermann, C., Karma, A. \& Li, Q. [2001] ``Phase-field simulations of dendritic crystal growth in a forced flow'', \textit{Phys. Rev. E} \textbf{63}, 061601-1-16.
\bibitem{Wang} Wang, S.-L., Sekerka, R. F., Wheeler, A. A., Murray, B. T., Coriell, S. R., Braun, R. J. \& McFadden, G. B. [1993] ``Thermodynamically-consistent phase-field models for solidification'', \textit{Physica D} \textbf{69}, 189-200.


\end{thebibliography}
\end{document}